\documentclass[aip,reprint,english,a4paper] {revtex4-1}

\usepackage{upgreek}
\usepackage{amsfonts}
\usepackage{amstext}
\usepackage{mathtools}
\usepackage{amssymb}
\usepackage{color}
\usepackage[normalem]{ulem}
\usepackage{amsmath} 
\usepackage{graphicx} 
\usepackage{bbold}

\usepackage{etoolbox}    
\usepackage{tikz}

\newrobustcmd*{\mysquare}[1]{\tikz{\filldraw[draw=#1,fill=#1] (0,0)
rectangle (0.1cm,0.1cm);}}

\newrobustcmd*{\mycircle}[1]{\tikz{\filldraw[draw=#1,fill=#1] (0,0) circle [radius=0.05cm];}}

\newcommand{\vx}{\boldsymbol{x}}
\newcommand{\vy}{\boldsymbol{y}}
\newcommand{\vf}{\boldsymbol{f}}
\newcommand{\mmass}{m}
\newcommand{\mG}{\Gamma}
\newcommand{\mK}{\mathcal{K}}
\newcommand{\vxi}{\boldsymbol{\xi}}

\newcommand{\SIhierarchyGamma}      {S1}
\newcommand{\SInucleationNoiseTest} {S2}

\newcommand{\SIAibACF}              {S4}
\newcommand{\SIAibWaitingTimes}     {S5}

\begin{document}

\author{Benjamin Lickert, Steffen Wolf and Gerhard Stock}
\email{stock@physik.uni-freiburg.de}
\affiliation{Biomolecular Dynamics, Institute of Physics, Albert Ludwigs
	University, 79104 Freiburg, Germany.}
\title{Data-driven Langevin modeling of nonequilibrium processes}
\date{\today}

\begin{abstract}
  Given nonstationary data from molecular dynamics simulations, a
  Markovian Langevin model is constructed that aims to reproduce the
  time evolution of the underlying process. While at equilibrium the
  free energy landscape is sampled, nonequilibrium processes can be
  associated with a biased energy landscape, which accounts for finite
  sampling effects and external driving. Extending the data-driven
  Langevin equation (dLE) approach [Phys.\ Rev.\ Lett.\ {\bf 115},
  050602 (2015)] to the modeling of nonequilibrium processes, an
  efficient way to calculate multidimensional Langevin fields is
  outlined.  The dLE is shown to correctly account for various
  nonequilibrium processes, including the enforced dissociation of
  sodium chloride in water, the pressure-jump induced nucleation of a
  liquid of hard spheres, and the conformational dynamics of a helical
  peptide sampled from nonstationary short trajectories.
\end{abstract}
\maketitle

%
%

\section{Introduction}

Classical molecular dynamics (MD) simulations may account for the
structure and dynamics of molecular systems in microscopic detail, but
become cumbersome with increasing system size.\cite{Berendsen07} To
drastically reduce the complexity, it is therefore often desirable to
represent the considered process in terms of a small set of collective
variables $\vx=\{x_i\}$, also referred to as order parameters or
reaction coordinates.\cite{Rohrdanz13,Peters16,Sittel18} Employing
projection operator methods, \cite{Grabert80,Kubo85, Zwanzig01} in
principle exact equations of motions for these variables, such as the
generalized Langevin equation (GLE), can be derived. If we assume a
time scale separation between the fast degrees of freedom (the bath)
and the slow collective variables $\vx$ (the system), we obtain a
Markovian Langevin equation (LE)
\begin{equation}\label{eq:LE}
\mmass \ddot{\vx}(t)= \vf(\vx)- \mG(\vx) \dot{\vx}(t)+\mK(\vx) \vxi(t),
\end{equation}
which describes the stochastic motion of a quasi-particle with mass
$\mmass$. Here the deterministic Newtonian force
$\vf=- \partial_{\vx} \Delta G(\vx)$ is given by the gradient of the
free energy landscape $\Delta G$, $\mG$ accounts for the friction
tensor, $\mK$ represents the diffusion tensor, and $\vxi(t)$
represents Gaussian-distributed white noise of zero mean. Friction and
diffusion in general depend on variable $\vx$ and are connected at
thermal equilibrium by the fluctuation-dissipation
theorem,\cite{Zwanzig01} $\mK\mK^T=2k_{\rm B}T\mG$.

\begin{figure}[ht!]
\includegraphics[width=0.40\textwidth]{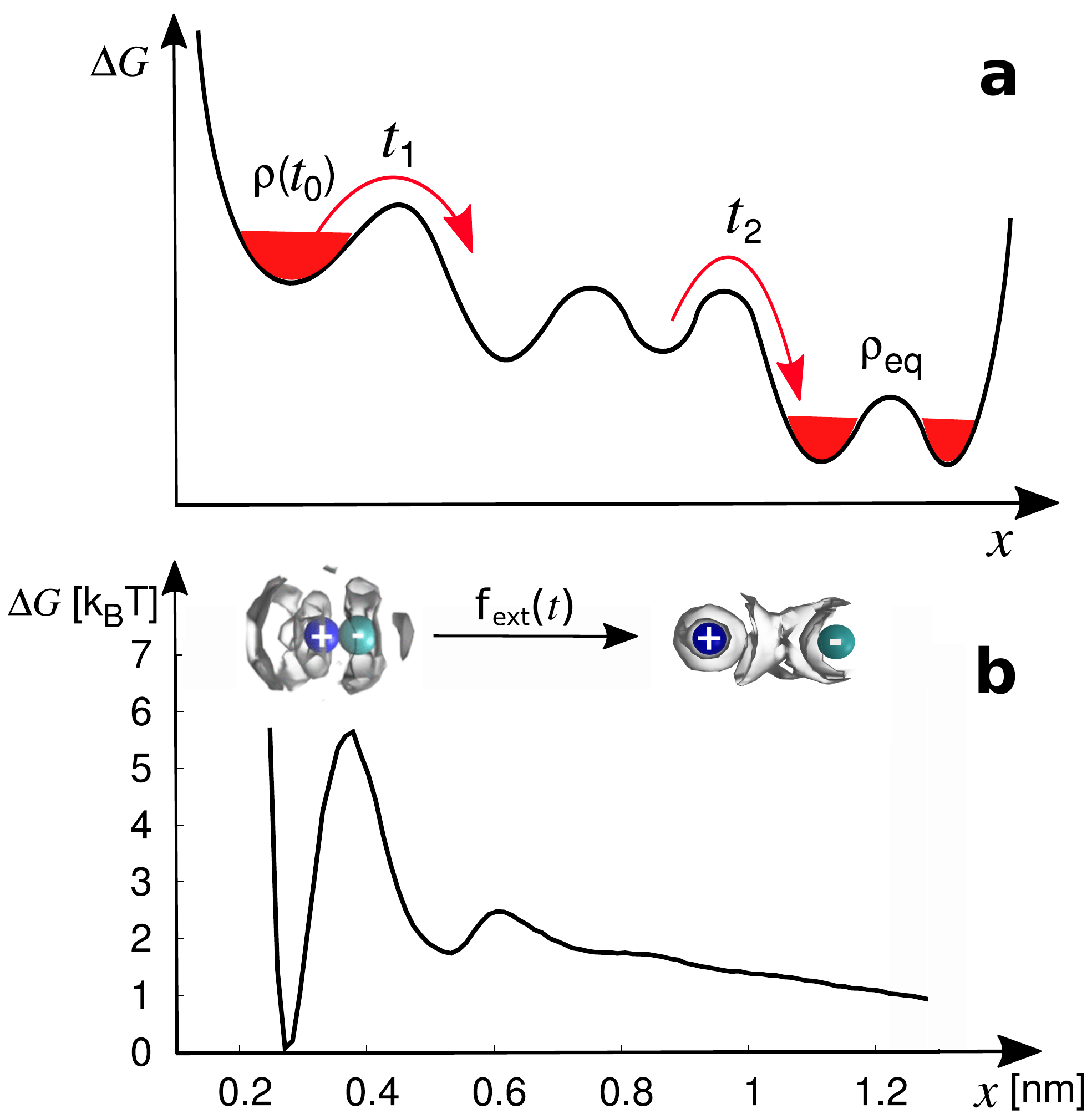}
\caption{
  Nonequilibrium processes considered. (a)
  Relaxation: Starting in a nonstationary state $\rho(t_0)$, the
  system evolves along the free energy landscape $\Delta G(x)$ towards
  its equilibrium state $\rho_{\rm eq}$. (b) External driving:
  Application of an external pulling force $f_{\rm ext}(t)$ along the
  interionic distance $x$ causes the dissociation of solvated NaCl
  along $\Delta G(x)$.}
\label{fig:neq} 
\end{figure}

While Langevin theory is well established, \cite{Grabert80,Kubo85,Zwanzig01} a
relatively little explored aspect concerns its application to
nonequilibrium processes. For one, we may start in a nonstationary
state $\rho(t_0)$ (e.g., resulting from a laser-induced
temperature-jump or photoexcitation) and study the relaxation of the
(otherwise not driven) system into its equilibrium state
$\rho_{\rm eq}$ (Fig.\ \ref{fig:neq}a). On the other hand, we may
consider a system driven by an external force $f_{\rm ext}(t)$, as is
the case, e.g., in atomic force microscopy experiments, where we
mechanically pull a molecular complex apart (Fig.\ \ref{fig:neq}b).
To account for such processes, the standard Langevin formalism can be
generalized in various ways. \cite{Hernandez99, McPhie01,
  Micheletti08,Kawai11,Meyer17,Meyer19,Cui18} Using projection
operator techniques, for example, several nonstationary versions of
the GLE have been derived.\cite{Kawai11,Meyer17,Meyer19}
Alternatively, we may consider a microscopic Hamiltonian of the full
system, and derive equations of motion for the collective
variables. \cite{Zwanzig73} This formulation can be readily extended
to describe nonequilibrium relaxation processes by considering
nonstationary initial conditions.\cite{Cui18} In the case of
external driving, we expect that in the linear response regime the
external force affects mainly the drift term $\vf(\vx)$ but neither
the friction $\mG$ nor the noise $\mK$ of the LE.\cite{Cui18}

Here we are concerned with the situation that we have nonstationary
data $\vx (t)$ (provided by an MD simulation or a time-dependent
experiment) for which we want to construct a dynamical model that
reproduces its time evolution. To this end, we need to determine the
underlying parameters or fields of the model from the MD
data.\cite{Straub87,Hummer05, Best06,
  Lange06b,Hinczewski10,Perez-Villa18, Wolf18,Paul19} For example,
when we employ the Markovian LE (\ref{eq:LE}), we want to calculate
the Langevin fields $\vf$, $\mG$ and $\mK$.  To evaluate
multidimensional Langevin fields from given MD data, we have developed
a data-driven Langevin equation (dLE) approach, which was successfully
applied to model the conformational dynamics of various peptides and
proteins at equilibrium.\cite{Hegger09,Schaudinnus13, Schaudinnus15,
  Schaudinnus16,Lickert20}

To extend the application of the dLE to the nonequilibrium regime, we derive the
necessary modifications of the dLE algorithm for both relaxation and
driven processes. In particular, we discuss the practical computation of
nonstationary Langevin fields and discuss the dependence of these
estimates on the sampling quality of the input data such as their
length. As first applications, we demonstrate that
the Markovian LE (\ref{eq:LE}) correctly accounts for two
nonequilibrium processes recently studied via a nonstationary GLE,
\cite{Meyer17,Meyer19} that is, the enforced dissociation of NaCl in
water\cite{Meyer20} and the pressure-jump induced nucleation and
growth process in a liquid of hard spheres.\cite{Meyer21} In both
cases, a simple Langevin model is constructed that is shown to reproduce
well the underlying MD reference data.
As a Langevin-based enhanced sampling strategy,\cite{Rzepiela14} we
finally consider the construction of a global Langevin model that
accounts for the overall kinetics of the system, obtained from massive parallel
computing of short MD trajectories that
are {\em per se} nonstationary. Modeling the conformational dynamics
of the helical peptide Aib$_9$ in five dimensions,\cite{Buchenberg15} we
compare the predictions of the resulting dLE to a recently established
Markov state model of the system.\cite{Biswas18} 
%

%
%
\section{Theory and Methods}
\subsection{Data-driven Langevin equation (dLE)} \label{sec:dLE} 
  
To briefly review the gist of the dLE
approach,\cite{Hegger09,Schaudinnus13, Schaudinnus15, Schaudinnus16,Lickert20} we
consider a discrete time series $\vx_n\equiv\vx(n\delta t)$ of the
dimensionless coordinate $\vx$ with time step $\delta t$. (For
clarity, we henceforth use a vector notation.) By discretizing the
time derivatives in Eq.\ (\ref{eq:LE}), i.e,
$\dot{\vx}_n\rightarrow (\vx_n-\vx_{n-1})/{\delta t}$, we obtain a
discrete version of the Langevin equation\cite{Schaudinnus15}
\begin{equation}\label{dLE}
\vx_{n+1}=\vx_n+\hat{\vf}_n-\hat{\mG}_n(\vx_n-\vx_{n-1})+\hat{\mK}_n\vxi_n ,
\end{equation}
where we introduced the dLE fields
\begin{equation}\label{dLEfields}
\begin{aligned}
    \hat{\vf}_n  &= \mmass^{-1}\delta t^2\vf(\vx_n),             &
    \hat{\mG}_n &= \mmass^{-1}\delta t\mG(\vx_n)-\mathbb{1},  \\  
    \hat{\mK}_n  &= \mmass^{-1}\delta t^{3/2}\mK(\vx_n),  &  \vxi_n &=
    \vxi(\vx_n)\sqrt{\delta t},
\end{aligned}
\end{equation}
accounting for drift $\hat{\vf}(\vx)$, friction $\hat{\mG}(\vx)$ and
noise amplitude $\hat{\mK}(\vx)$ of the Langevin model.

Equation \eqref{dLE} can be used in two ways. In case that the dLE
fields are given, it may be integrated to yield the time evolution of
coordinate $\vx_n$. On the other hand, if the time series $\vx_n$ is
given, we face the inverse problem and need to solve for the dLE
fields defined by Eq.\ (\ref{dLEfields}). Because of the stochastic
nature of the noise $\vxi_n$, we can not determine these fields
directly from the individual data points.\cite{Hegger09} Hence we
introduce a local average of some observable $g(\vx)$
\begin{equation}
\langle g(\vx)\rangle=\frac{1}{k}\sum_{m}g(\vx_m),
\label{knextneighav}
\end{equation}
where the sum includes the $k$ nearest neighbors of reference point
$\vx$ which occur in the input time series. The neighborhood size $k$
needs to be small enough to allow for local (i.e.,
coordinate-dependent) averaging, but at the same time it should be
large enough to achieve statistical convergence.\cite{Schaudinnus13}
As a rule of thumb, a value of $k\approx 200$ performs well if $\gtrsim 10^6$
data points are available, and is therefore used in all cases below.

By exploiting the white noise properties $\langle \vxi_n\rangle=0$ and
$\langle \vxi_i\vxi_j\rangle=\delta_{ij}$, we can derive from Eq.\
(\ref{dLE}) explicit expressions for the dLE
fields\cite{Schaudinnus15}
\begin{align}
\hat{\mG}_n  &= -C(\Delta \vx_{n+1},\Delta \vx_{n})C^{-1}(\Delta
\vx_{n},\Delta \vx_{n}), \label{eq:Gamma}\\ 
    \hat{\vf}_n &= \langle\Delta \vx_{n+1}\rangle
    +\hat{\mG}_n\langle\Delta \vx_{n}\rangle,\label{eq:drift}\\  
    \hat{\mK}_n\hat{\mK}^T_n  &= C(\Delta \vx_{n+1},\Delta
    \vx_{n+1})\!-\!\hat{\mG}_nC(\Delta \vx_{n},\Delta
                                \vx_{n+1}) \label{eq:noise}, 
\end{align}
where we introduced the displacement
$\Delta \vx_n\!=\!\vx_n\!-\!\vx_{n-1}$ and the covariance
$C(\vx,\vy)=\langle\vx\vy^T\rangle-
\langle\vx\rangle\langle\vy^T\rangle$, with averages as in Eq.\
(\ref{knextneighav}). In a last step, a Cholesky decomposition of
$\hat{\mK}_n\hat{\mK}^T_n$ is used to calculate the noise amplitude
matrix $\hat{\mK}_n$.\cite{Schaudinnus13} 
Equations (\ref{dLE}) -- (\ref{eq:noise}) constitute the working
equations of the dLE. That is, by calculating dLE fields
$\hat{\vf}_n$, $\hat{\mG}_n$ and $\hat{\mK}_n$ from the input data
$\vx_{n}$, we propagate Eq.\ (\ref{dLE}) to obtain a dLE trajectory,
which then can be employed to predict statistical and dynamical
properties of the system.

Since the dLE is based on a Markovian approximation, it requires a
propagation time step $\delta t$ that is longer than the memory time
of the system, while $\delta t$ simultaneously needs to
be short enough to resolve the dynamics
of the system. Choosing $\delta t$ to meet the second
condition, we may test the validity of the resulting dLE model by
imagining that the input trajectory $\vx_n$ was generated by the
propagation of Eq.\ (\ref{dLE}). Analyzing the properties of the
resulting noise trajectory
\begin{equation}\label{dLEnoise}
\vxi_n=\hat{\mK}_n^{-1}(\vx_{n+1}-\vx_n-
\hat{\vf}_n+\hat{\mG}_n(\vx_n-\vx_{n-1})),
\end{equation}
we can check if the noise indeed fulfills the criteria of white noise
such as zero mean and $\delta$-correlation. \cite{Schaudinnus13}
Failing this test indicates that the time step is too short for the
presumed Markov approximation of the dLE model. As a remedy, we
recently proposed to rescale the friction and the noise tensor of the
dLE via $\mG\rightarrow S\mG S^T$ and $\mK\rightarrow S\mK$, in order
to obtain the correct long-time behavior of the
system.\cite{Lickert20} To determine the scaling factor $S$, we require
that the resulting dLE model reproduces the correct initial decay of
the position autocorrelation function. 

To facilitate the treatment of multidimensional systems, the dLE
fields are calculated ``on the fly'', i.e., at every propagation step
of the dLE trajectory.\cite{Hegger09} The drawback of this strategy is
that the local averaging in Eq.\ (\ref{knextneighav}) requires the
determination of the $k$ nearest neighbors of all $N$ data points at
every dLE step. Besides using an efficient box-assisted search
\cite{Grassberger90} (which scales $\propto N \ln N$), we recently
introduced a ``pre-averaging'' approach which exploits the fact that
the dLE fields are estimated from local neighborhoods, and allows for
a massive reduction of data points (say, from $10^7$ input points to
$10^5$ averaged points) that need to be scanned in every dLE
step (see SI methods for details).\cite{Lickert20}

Let us finally note two features of the dLE that are particularly
valuable for its application to general nonequilibrium
processes. First off, the calculation of dLE fields in Eqs.\
(\ref{dLE}) -- (\ref{eq:noise}) in principle requires only short
trajectory pieces (i.e., at least three successive time steps). This
local information does not require global equilibrium data, but can be
readily obtained from short nonstationary data.\cite{Rzepiela14}
Secondly, we note that the dLE of time series $\vx(t)$ in Eq.\
(\ref{dLE}) does not involve the mass tensor $\mmass$, because it is
included in the definition of the dLE fields. Hence, we are not
restricted to the modeling of the motion of particles, but can use
Eq.\ (\ref{dLE}) to model the stochastic time evolution of an
arbitrary phase-space function $\boldsymbol{A}(t)$.

%
%
\subsection{dLE modeling of relaxation processes} \label{sec:relax}

To apply the dLE formulation to nonstationary data, we first focus on
the case shown in Fig.\ \ref{fig:neq}a, where the system is prepared
at time $t_0\!=\!0$ in a nonstationary state $\rho(t_0)$ and for
$t \!>\! t_0$ relaxes towards its equilibrium state $\rho_{\rm eq}$.
While the initial state of the process is by design nonstationary, the
definition of the system (e.g., via a microscopic system-bath
Hamiltonian or an MD force field) is the same as for an equilibrium
process and therefore associated with the free energy landscape
$\Delta G(\vx)$. Since the Langevin fields $\vf$, $\mG$ and $\mK$
derive from this Hamiltonian,\cite{Zwanzig01} they should be the same
as well. In principle we can therefore again construct an appropriate
Langevin model from a long equilibrium trajectory, and use local
averages [Eq.\ (\ref{knextneighav})] to calculate the dLE fields
$\hat{\vf}_n$, $\hat{\mG}_n$ and $\hat{\mK}_n$.

In practice, though, equilibrium MD data would hardly sample the
initial high-energy regions of the process (Fig.\ \ref{fig:neq}a),
which are of interest when we study the time evolution of the
relaxation. Similar to experiment, we therefore perform an ensemble
average over numerous nonequilibrium trajectories
$\vx^{(r)}(t_n) \equiv \vx_n^{(r)}$ ($r = 1,\ldots,N_{\rm traj}$) of
some length $t_{\rm max}$, and calculate local averages of the dLE
fields from this data. As a consequence, the accuracy of the dLE field
estimation will depend on the sampling parameters $N_{\rm traj}$ and
$t_{\rm max}$. For example, in the case of a short trajectory length
$t_{\rm max}$, the model may only account for the initial dynamics at
times $t\lesssim t_1$, but can not describe its further relaxation to
equilibrium for $t\gtrsim t_2$ (Fig.\ \ref{fig:neq}a).

To discuss the effects of a finite sampling time $t_{\rm max}$, we
consider the resulting drift field of the dLE, given by
$\vf(\vx)=- \partial_{\vx} \Delta G(\vx)$. At equilibrium,
$\Delta G(\vx)$ represents the free energy landscape,
$\Delta G(\vx)=-k_{\rm B}T \ln P(\vx)$ with equilibrium distribution
$P(\vx)$. When we estimate the potential from $N_{\rm traj}$
nonequilibrium trajectories of length $t_{\rm max}$, we
correspondingly obtain a ``biased energy landscape'' \cite{Post19}
$\Delta{\cal G} (\vx)$ that depends on the distribution
${\cal P}(\vx)\propto\int dt\rho(\vx,t)$ sampled within
$0 \le t \le t_{\rm max}$,
\begin{equation} \label{eq:PMF}
\Delta{\cal G} (\vx) =-k_{\rm B}T \ln {\cal P}(\vx) ,
\end{equation}
and approaches the equilibrium free energy landscape only for
$t_{\rm max}\!\rightarrow\!\infty$. Limited
sampling in particular means that $\Delta{\cal G} (\vx)$ may not cover
all parts of the equilibrium energy landscape $\Delta G(\vx)$, see
Sec.\ \ref{sec:1dHierarch}. Note that the term ``biased'' here refers 
to the nonstationary initial conditions, and should not be confused
with biased potentials in enhanced sampling techniques.\cite{Berendsen07} 

We note that both, $\Delta G(\vx)$ and $\Delta{\cal G} (\vx)$,
represent a global observable of the system. The drift field
$\vf\!=\!-\partial_{\vx} \Delta{\cal G}$, on the other hand, measures the
slope of the energy landscape at position $\vx$. It can be therefore
estimated by the local average in Eq.\ (\ref{knextneighav}) via a sum
over the $k$ nearest neighbors of $\vx$. Because the dLE fields are
generally based on local averages, the same argument hold also for the
friction $\hat \mG_n \equiv \hat \mG(\vx(t_n))$ [Eq.\
(\ref{eq:Gamma})] and the noise $\hat \mK_n \equiv \hat \mK(\vx(t_n))$
[Eq.\ (\ref{eq:noise})].

To construct a dLE model of nonequilibrium relaxation, we therefore
(i) calculate the dLE fields for the given nonstationary data and (ii)
perform dLE runs using initial conditions of the relaxation process.
While no additional assumptions are involved (compared to an
equilibrium dLE), the nonstationary dLE model depends on the sampling
achieved by the input data, in particular on their length (see Sec.\
\ref{sec:1dHierarch}).

%
%
\subsection{dLE modeling of external driving} 

Unlike to the modeling of relaxation processes, where the dLE fields
are in principle the same as at equilibrium, external driving via some
force $\vf_{\rm ext}(\vx,t)$ will change these fields. First of all,
the deterministic drift term $\vf(\vx)$ in Eq.\ (\ref{dLE}) needs to
be complemented by the external perturbation $\vf_{\rm ext}(\vx,t)$.
In a microscopic description, such as an all-atom MD simulation, this
is all to be done - external forces can simply be added to the internal
forces. In a reduced description such as a Langevin model, however,
external forces may in principle also affect the friction $\mG$ and
the noise $\mK$ of the model.\cite{Meyer17,Meyer19,Daldrop17} 
In the linear-response regime, we nevertheless expect that
nonequilibrium and equilibrium dynamics exhibit the same friction and
noise. According to Onsager's regression hypothesis,\cite{Chandler87}
this is because in this regime nonequilibrium perturbations are
governed by the same laws as equilibrium fluctuations. For example,
when we derive LE (\ref{eq:LE}) from a system-bath ansatz where a
general system is linearly coupled to a harmonic bath,\cite{Zwanzig73}
we find that external and internal forces are simply
added.\cite{Cui18}

Given a dLE model constructed from equilibrium data and an external
potential $V_{\rm ext}(\vx,t)$ that affects a linear response of the
system, we thus obtain a dLE for the driven nonequilibrium processes
by replacing the free energy landscape $\Delta G(\vx)$ of the equilibrium
model by the biased energy landscape
\begin{equation} \label{eq:PMF2}
\Delta{\cal G} (\vx, t) = \Delta G(\vx) + V_{\rm ext}(\vx,t).
\end{equation}
As a simple example, we consider an atomic force microscopy experiment
that causes the dissociation of a diatomic molecule (Fig.\
\ref{fig:neq}b). The one-dimensional pulling can be modeled via
a harmonic spring, \cite{Grubmueller96,Isralewitz01,Park04} 
\begin{equation} \label{eq:drive}
V_{\rm ext}(x,t) = -\frac{C}{2} [x(t) - (x_0+vt)]^2,
\end{equation}
where $C$ is the spring constant and $v$ the pulling velocity. As a
consequence, the system is gradually pulled along the free energy
landscape $\Delta G(x)$ in Fig.\ \ref{fig:neq}b. That is, the biased
energy landscape $\Delta{\cal G}$ corresponds to the total
deterministic potential experienced by the system at time $t$.  This
simple picture is expected to be valid for most biomolecular
applications taking place close to equilibrium. Considering the
enforced dissociation of NaCl discussed below, for example, we found
linear response behavior for pulling velocities up to $\sim 10\,$ m/s,
at which the pulling approaches the picosecond timescale of the
reordering of the solvation shells.\cite{Wolf18}

To go beyond the linear response or to construct a dLE directly from
data of a driven system, we need to calculate explicitly
time-dependent dLE fields from the nonequilibrium data. Hence the
local average of function $g(\vx)$ in Eq.\ (\ref{knextneighav}) has to
be replaced by a time-dependent average $\langle g(\vx,t)\rangle$.
Assuming that the nonequilibrium process is sampled by $N_{\rm traj}$
trajectories $\vx_n^{(r)},$ this ensemble average can be written as
\begin{equation} \label{timeDepAv}
\langle g(\vx,t_n)\rangle = \frac{1}{N_{\rm traj}}\sum_{r=1}^{N_{\rm
    traj}} g(\vx_n^{(r)}) \, \delta_k (\vx -\vx_n^{(r)}),
\end{equation}
where $\delta_k (\vx -\vx_n^{(r)})$ defines a boxing function that is equal
to $1/k$ for the $k$ nearest neighbors of $\vx$ and zero otherwise. 

%
%
\subsection{Comparison to the nonstationary GLE} \label{sec:comp} 

The Markovian LE (\ref{eq:LE}) is based on a time scale separation
between the fast bath degrees of freedom and the slow collective
variable $\vx$. On the other hand, various generalized LEs (GLEs) have
been proposed that are not limited to a Markov-type approximation. To
describe nonequilibrium processes, in particular, Schilling and
coworkers recently employed a time-dependent projection operator
formalism to derive for some variable $A(t)$ the nonstationary GLE
\cite{Meyer17,Meyer19}
\begin{equation}\label{eq:nsGLE}
 \dot{A}(t)= \omega (t) A(t) + \int_0^t d\tau 
K(t,\tau) A(\tau) + \eta(t),
\end{equation}
where memory kernel $K$ and noise $\eta$ are related by a generalized
fluctuation-dissipation theorem. Due to the Mori-type projection
employed, the equation exhibits a linear drift term $\omega (t)A(t)$
that vanishes for a mean-free variable $A(t)$.

It is instructive to compare this formulation to our Markovian
Langevin approach, also because we adopt below two model
problems\cite{Meyer20,Meyer21} that were previously studied with this
GLE. In these studies, the GLE analysis resulted in slowly decaying
memory kernels, which appears to question the applicability of a
Markovian Langevin model such as the dLE.  However, whether a
timescale separation is given depends on the system-bath partitioning
of the problem. As we show in the following, the LE (\ref{eq:LE}) and
the GLE (\ref{eq:nsGLE}) are based on different partitionings, where
the former may be valid, although the latter
exhibits non-Markovian behavior.

Following Zwanzig,\cite{Zwanzig01} we explain this
difference for the case of a one-dimensional system variable $x(t)$.
To begin, we rewrite the second-order differential equation
(\ref{eq:LE}) for $x(t)$ in terms of two first-order equations,
\begin{eqnarray}
\dot{x}(t) &=& p(t)/\mmass , \label{eq:LEx}\\
\dot p (t) &=& f(x) - \mG p(t)+\mK \xi(t),   \label{eq:LEp}
\end{eqnarray}
for the phase space variables $x(t)$ and $p(t)$. Integrating Eq.\
(\ref{eq:LEp}), we obtain
\begin{align}
\begin{split}
p(t)=  & \int_0^t \mathrm{d}\tau\, {\mathrm e}^{-\mG(t-\tau)/\mmass}f(x(\tau)) \\
&+ \int_0^t \mathrm{d}\tau\, {\mathrm e}^{-\mG(t-\tau)/\mmass} \mK \xi(\tau),
\end{split}
\end{align}
where we assumed that $p(0)\!=\!0$. Insertion in Eq.\
(\ref{eq:LEx}) yields the desired GLE for $x(t)$,
\begin{eqnarray} \label{eq:GLE}
\dot{x}(t) =  \int_0^t \mathrm{d}\tau K(t,\tau) x(\tau) + \eta(t).
\end{eqnarray}
It contains a two-time memory kernel
\begin{equation}  \label{eq:Ktt}
K(t,\tau) = \frac{1}{\mmass}\, {\mathrm e}^{-\mG(t-\tau)/\mmass} 
 \,\frac{f(x(\tau))}{x(\tau)},
\end{equation}
which in the case of a linear force
(${f(x(\tau))}/{x(\tau)}\!=$const.) reduces to a simple exponential
function that depends only on the time difference $t\!-\!\tau$. The
function $\eta(t)$ represents the corresponding colored noise. Note
that the GLE (\ref{eq:GLE}) for $x(t)$ is equivalent to the Markovian
LEs (\ref{eq:LEx}) and (\ref{eq:LEp}) for $x(t)$ and $p(t)$, that is,
both formulations describe the same physics. Non-Markovian features of
the GLE such as a slowly decaying memory kernel (\ref{eq:Ktt}) may
therefore reflect a slowly varying force, which
arises from a multistate energy landscape with rarely occurring
transitions. 

Because Eq.\ (\ref{eq:GLE}) is of similar form as the nonstationary
GLE (\ref{eq:nsGLE}), the above argument holds also in this
case. Nonetheless, the nonstationary GLE represents an exact
formulation,\cite{Meyer17,Meyer19} that is, for any nonequilibrium
process a GLE (\ref{eq:nsGLE}) can be defined via the functions
$\omega(t)$, $K(t,\tau)$ and $\eta(t)$ that do not depend on
coordinate $x$.
Employing the dLE, on the other hand, we need to assume that a
Markovian LE model exists for the considered process. As a trade-off
we obtain a physically intuitive model, where the biased energy
landscape $\Delta{\cal G}$ accounts for the main features of the
considered process, and the friction $\mG$ and the noise $\mK$
describe in a transparent way the effects of the bath.

%
%
\section{Results and discussion} \label{sec:Results}
\subsection{Hierarchical energy landscape} \label{sec:1dHierarch} 

\begin{figure}[ht!]
\includegraphics[width=0.4\textwidth]{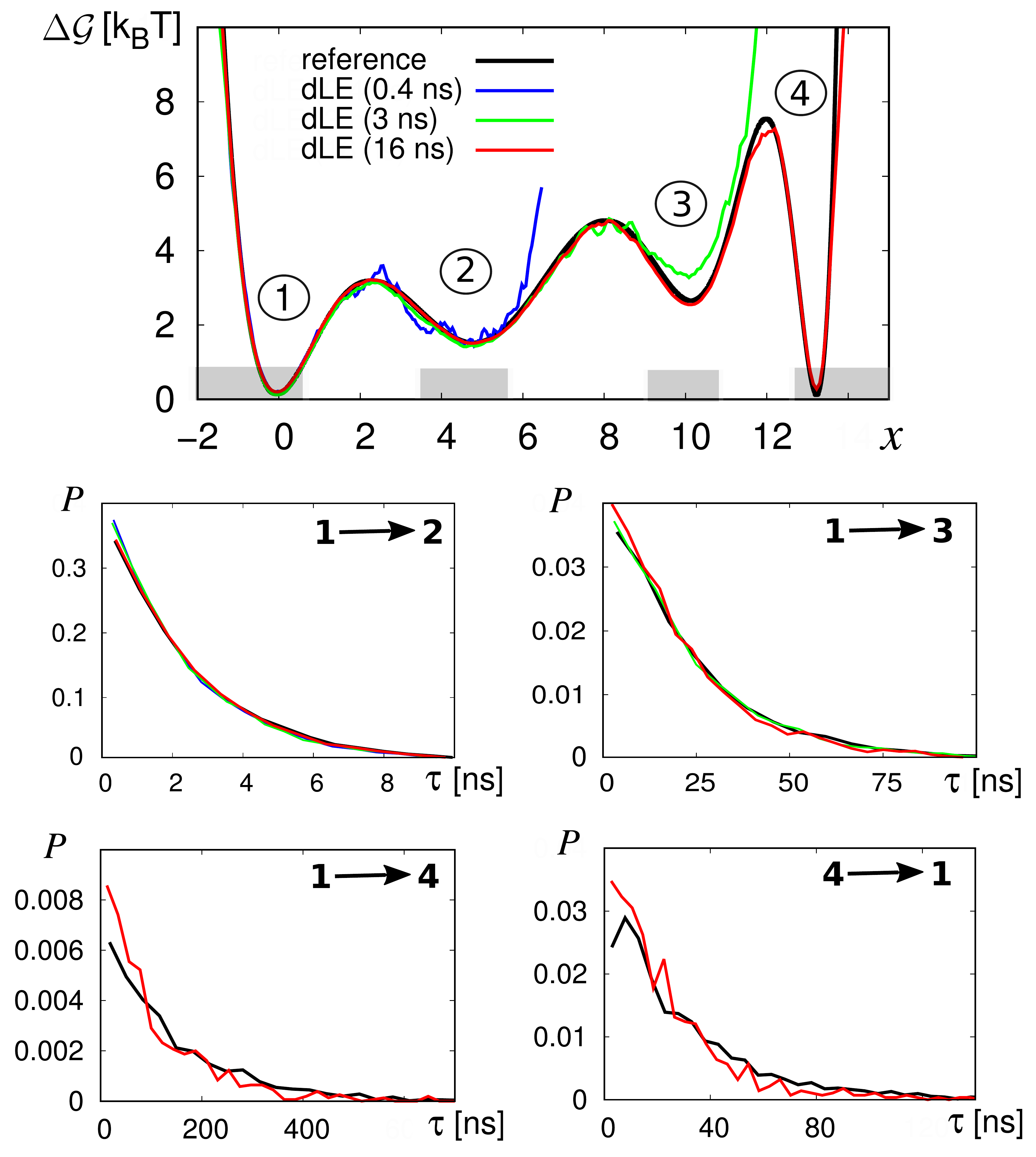}
\caption{
  (Top) Hierarchical free energy landscape (black), which consists of
  states ${\bf 1}-{\bf 4}$ connected via energy barriers of similar
  height. Gray regions at the bottom define cores of the states used
  to calculate waiting times. Starting in state {\bf 1}, sets of
  trajectories of lengths $t_{\rm max} = 0.4\,$ns, 3\,ns, and 16\,ns
  are generated, which result in nonequilibrium energy landscapes
  $\Delta{\cal G} (x)$ [Eq.\ (\ref{eq:PMF})] reaching state {\bf 2}
  (blue), {\bf 3} (green), and {\bf 4} (red), respectively. (Bottom)
  Distributions of waiting times $\tau_{i\rightarrow j}$ of
  transitions {i}$\rightarrow${j} as indicated. Results from dLE
  models using input data of length 0.4\,ns (blue), 3\,ns (green), and
  16\,ns (red) are compared to reference data (black).}
\label{fig:1dmodel} 
\end{figure}

We begin with the first nonequilibrium scenario where the system is
initially prepared in a nonstationary state and subsequently relaxes
into its equilibrium state (Fig.\ \ref{fig:neq}a).  As discussed in
Sec.\ \ref{sec:relax}, the dLE modeling of nonequilibrium relaxation
processes may sensitively depend on the length $t_{\rm max}$ of the
input data. This is particularly true for biomolecular systems that
are characterized by a hierarchical energy landscape, which gives rise
to coupled processes on timescales ranging from, say pico- to
milliseconds. \cite{Frauenfelder91, Henzler-Wildman07,Buchenberg15} To
demonstrate the effect of finite data length on a dLE model, we
consider a one-dimensional free energy landscape with three
consecutive energy barriers of similar height, see Fig.\
\ref{fig:1dmodel}. The model aims to mimic photo- or ligand-induced
conformational transitions in proteins,\cite{Buchenberg17,Bozovic20}
where at time $t\!=\!0$ the system is prepared in the
nonstationary state {\bf 1}, evolves via intermediate states {\bf 2}
and {\bf 3}, and finally reaches the second low-energy state {\bf
  4}. Due to the hierarchical shape of the free energy landscape,
transition {\bf 1}$\rightarrow${\bf 2} occurs much more rapidly than
transition {\bf 1}$\rightarrow${\bf 3} or even {\bf
  1}$\rightarrow${\bf 4}, because the latter require the previous
transitions as a prerequisite step.

Using this energy landscape, a mass of $\mmass=26\,$u, constant
friction ($\mG\!=\!130\,$kJ ps/(mol nm$^2$)) and noise amplitude
($\mK\!=\!25\,$kJ ps$^{1/2}$/(mol~nm)), a temperature $T=300\,$K, and
a time step $\delta t=0.04$ ps, we run $20\!\times\! 16 \,\mu$s-long
Langevin trajectories as reference data, each trajectory starting at
$x=0$. Here we are interested in the waiting times of transitions
between states $i$ and $j$,\cite{Gernert14} i.e., the time spend in
state $i$ and possible subsequent intermediate states before going to
state $j$. Figure \ref{fig:1dmodel} shows the resulting waiting time
distributions, Table \ref{Tab1} lists the corresponding mean values
$\tau_{i\rightarrow j}$. We find that
$\tau_{1\rightarrow 2} \!=\!2.5\,$ns is about one order of magnitude
shorter than $\tau_{1\rightarrow 3}\!=\!23\,$ns, which again is almost
an order of magnitude shorter than
$\tau_{1\rightarrow 4}\!=\!143\,$ns. Back transitions from states {\bf
  2}, {\bf 3} and {\bf 4} to state {\bf 1} occur with waiting times of
$2.4\,$ns, $8\,$ns, $30\,$ns, respectively. The resulting
nonequilibrium energy landscape $\Delta{\cal G} (x)$ [Eq.\
(\ref{eq:PMF})] is indistinguishable from the (analytically given)
free energy landscape in Fig.\ \ref{fig:1dmodel}.

Employing these simulation results as input data, we now construct a
dLE of the hierarchical dynamics. Since the dLE is by design clearly
suited to reproduce data obtained by a Markovian LE, we can focus on
the effect of limited input data length $t_{\rm max}$ on the resulting
model. Guided by the waiting time distributions shown in Fig.\
\ref{fig:1dmodel}, we construct three input data sets consisting of
$10^2$ LE trajectories each starting in state {\bf 1} with
$t_{\rm max} = 0.4\,$ns, 3\,ns, and 16\,ns. These times are just long
enough to catch about ten transitions from state {\bf 1} to state {\bf
  2}, {\bf 3} and {\bf 4}, respectively. Using a time step of
$\delta t\!=\!0.04$ ps and a pre-averaging of the input data sets to
$10^4$ points (see SI Methods), we run for each input data set
$10\!\times\! 10\,\mu$s-long dLE simulations, which means that we
sample the different transitions at least $10^3$ times.
The effect of the finite input data length is clearly shown by the
resulting dLE nonequilibrium energy landscapes $\Delta{\cal G} (x)$
which, depending on the chosen value for $t_{\rm max}$, cover the
first two, three or all four states (see Fig.\ \ref{fig:1dmodel}). In a
similar way, the dLE reproduces the friction $\mG(x)\!=\,$const.\ and
the noise amplitude $\mK(x)\!=\,$const.\ of the model (Fig.\
\SIhierarchyGamma).

Considering transition {\bf 1}$\rightarrow ${\bf 2}, we find that a
dLE constructed from $0.4\,$ns-long data predicts a mean waiting time
$\tau_{1\rightarrow 2}\!=\! 2.4\,$ns that is about 4\% shorter than
the reference (Tab.\ \ref{Tab1}). The corresponding distribution in
Fig.\ \ref{fig:1dmodel} reveals that this effect is mainly caused by
an overestimation of fast transitions. Increasing the input data
length, distribution and mean value converge to the reference
data. Similar results are also found for the {\bf 1}$\rightarrow ${\bf
  3} transition.
Transitions involving the low-energy state {\bf 4} show somewhat
larger deviations. Using input data of 16\,ns length, the waiting
times for the {\bf 1}$\leftrightarrow ${\bf 4} transition are
underestimated by about 20\,\%. Again we find that short input data
bias towards short waiting times.

\begin{table}[h]
\centering
\begin{tabular}{l|c|c|c|c}
{data} & $\tau_{\text{1}\rightarrow\text{2}}$ &
$\tau_{\text{1}\rightarrow\text{3}}$ &
$\tau_{\text{1}\rightarrow\text{4}}$ &
$\tau_{\text{4}\rightarrow\text{1}}$ \\
    \hline
    reference & $2.5\pm0.01$ & $23\pm0.2$ & $143\pm3.3$ & $30\pm0.7$ \\
    \hline
    dLE (0.4\,ns) & $2.4\pm0.02$ & -- & -- & -- \\
    \hline
    dLE (3\,ns) & $2.4\pm0.01$ & $22\pm0.3$ & -- & -- \\
    \hline
    dLE (16\,ns) & $2.5\pm0.02$ & $21\pm0.3$ & $110\pm4.3$ &$24\pm0.9$ \\
\end{tabular}
\caption{
  Average waiting times (in ns) of the transitions $i\rightarrow j$ of
  the hierarchical energy landscape in Fig.\ \ref{fig:1dmodel},
  obtained for the reference data and for dLE models with different
  length $t_{\rm max}$ of the input data (given in
  parenthesis). Errors are calculated as standard deviations of the
  mean.}
\label{Tab1}
\end{table} 

To summarize, while the input data for a dLE model obviously need to
be long enough to reach a specific conformational state at all, it is
noteworthy that only a few events are sufficient for the dLE to
qualitatively predict the waiting times to these state.

%
%
\subsection{Enforced ion dissociation of NaCl in water}

As an example of an externally driven process, we now consider the
dissociation of solvated sodium chloride, which has served as a
versatile model problem to test the validity of various Langevin
formulations. This includes a dLE model constructed from equilibrium
MD data, \cite{Lickert20} a Markovian LE obtained from
dissipation-corrected targeted MD calculations of the free energy and
the friction,\cite{Wolf20} and the nonstationary GLE constructed from
pulling simulations. \cite{Meyer20}
Here we adopt pulling simulations of NaCl along the interionic
distance $x$ with pulling velocity $v\!\!=\!\! 10\,$m/s, using the
external force in Eq.\ (\ref{timeDepAv}) and an ensemble average over
1000 nonequilibrium trajectories.\cite{Meyer20} This MD data serve as
a reference to study the applicability of a Markovian Langevin
description of this problem. As detailed in Ref.\
\onlinecite{Wolf18}, all MD simulations employed
Gromacs v2018 (Ref.\ \onlinecite{Abraham15}) in a CPU/GPU hybrid
implementation, using the Amber99SB* force field\cite{Hornak06,Best09}
and the TIP3P water model.\cite{Jorgensen83}

\begin{figure*}[ht!]
\includegraphics[width=0.8\textwidth]{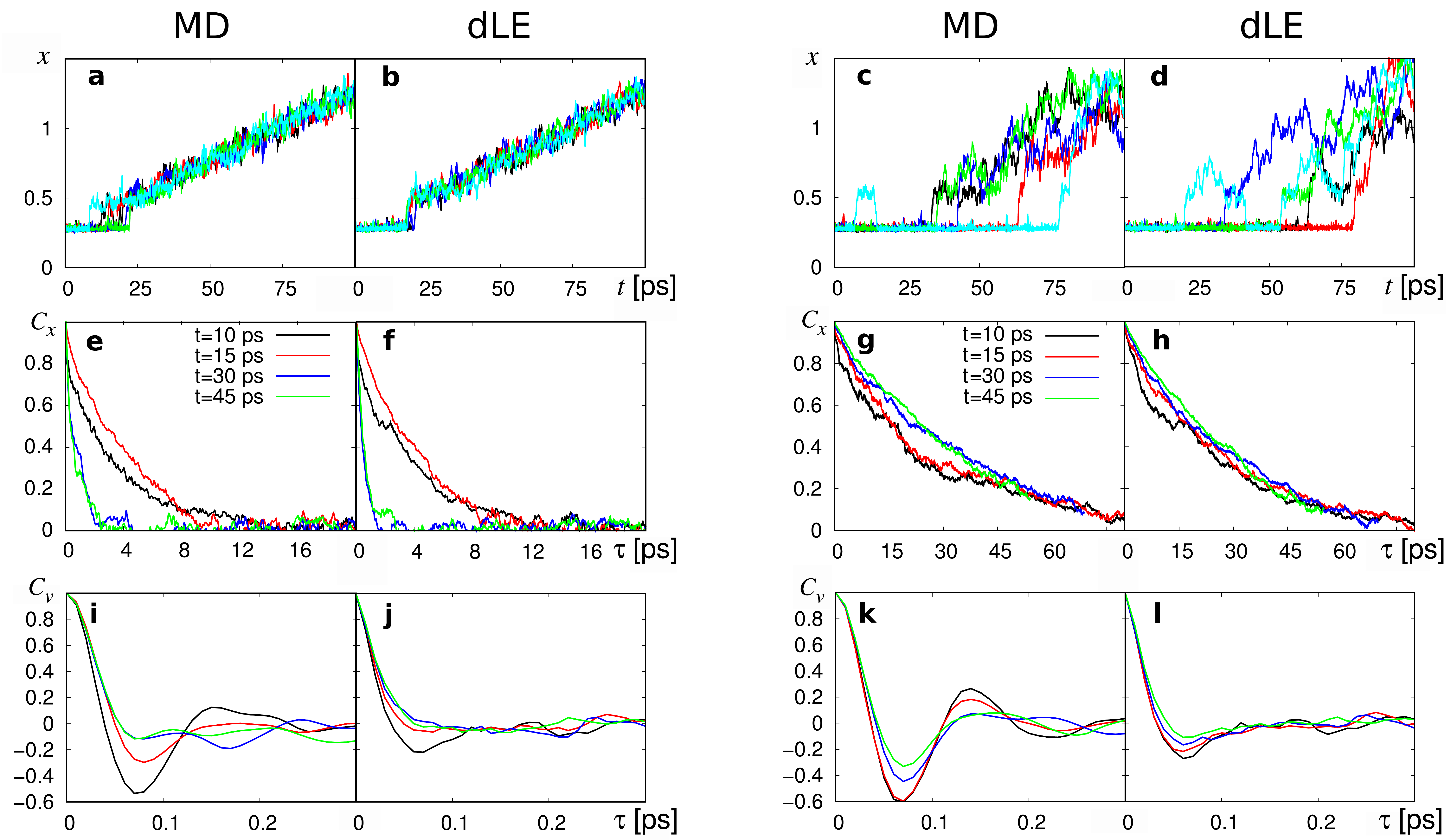}
\caption{
  Enforced dissociation of solvated NaCl. Comparing results from
  reference MD simulations and dLE model, we show (top) five sample
  trajectories, (middle) the position autocorrelation function $C_x$,
  and (bottom) the velocity autocorrelation function $C_v$, obtained
  for a strong spring (left) and a weak spring (right).}
\vspace*{-6mm}
\label{fig:NaCl} 
\end{figure*}

To illustrate the dissociation process, Fig.\ \ref{fig:neq}b shows the
free energy profile of NaCl along its interionic distance $x$, whose
main maximum at $x \approx 0.4$~nm corresponds to the
binding-unbinding transition of the two ions. The second smaller
maximum at $x \approx 0.6$~nm reflects the transition from a common to
two separate hydration shells.\cite{Mullen14}
Pulling NaCl apart using a strong ($C\!=\!1000\,$ kJ/(mol nm$^2$)) and
a soft ($C\!=\!100\,$ kJ/(mol nm$^2$)) spring, we obtain representative
trajectories $x(t)$ as shown in Figs.\ \ref{fig:NaCl}a,c,
respectively. For a strong spring, the system remains in the bound
state for $t \lesssim 20$~ps, before it abruptly moves over the main
barrier to the unbound state, where the Na-Cl distance $x(t)$ grows
linearly with time. The transition time of the individual trajectories
exhibits a narrow distribution with a width of the order of 10 ps. In
the weak spring simulations, on the other hand, the transition time
distribution is rather broad, such that the individual trajectories
seem to cross the barrier at random times.

To account for the statistical properties of the nonstationary time
series $x(t)$, we introduce the mean-free variable \cite{Meyer20}
\begin{equation}
\delta x(t) = \frac{x(t) - \langle x(t)\rangle}{\langle (x(t) - \langle
  x(t)\rangle)^2\rangle^{1/2}} \, ,
\end{equation}
which removes the systematic drift due to the pulling.
The autocorrelation function of this variable at time $t$,
\begin{equation}
C_x(t,t\!+\!\tau) = \langle \delta x(t)\delta x(t\!+\!\tau) \rangle
\end{equation}
drawn with respect to the delay time $\tau$ is shown in Figs.\
\ref{fig:NaCl}e,g for strong and soft springs, respectively. In the
former case, the autocorrelation decays on a timescale of 1~ps, except
when the system is at the barrier ($t = 10-15$~ps), where the decay is
significantly slower ($\sim 5$~ps). While the fast decay
of 1~ps can be associated with hydration shell dynamics around the
ions (which drives the ion dissociation), \cite{Wolf18} the slow decay is caused by forward- and backward
crossing events over the main free energy barrier of the system. For a weak spring, $C_x$ generally
decays slower ($\sim 20$~ps) for all times $t$.
		
We now turn to the Langevin modeling of the enforced dissociation of
NaCl. To this end, we adopt a previously described\cite{Lickert20} dLE
model constructed from equilibrium data at $300\,$K, which reproduces the MD
dissociation time (130 ps) and association time (850 ps) within an
error of a few percent. The model uses a sufficiently short
propagation time step to resolve the dynamics
($\delta t \!=\! 10\,$fs), the reduced mass of NaCl
($\mmass \!=\! 13.88$ u), a drift field corresponding to the free
energy profile shown in Fig.\ \ref{fig:neq}b, and constant friction
($\Gamma\!=\!594\,$kJ ps/(mol nm$^2$)) and noise amplitude
(${\cal K}\!=\!54\,$kJ ps$^{1/2}$/(mol~nm)). Adding the pulling force
[Eq.\ (\ref{timeDepAv})] to this dLE model, we wish to investigate to
what extent the resulting Langevin description can account for the MD
reference results.

By comparing the Langevin trajectories to the corresponding MD
results, Figs.\ \ref{fig:NaCl}a-d reveal that the Markovian LE mimics
the MD data quite closely. The same holds for the the position
autocorrelation function $C(t,t\!-\!\tau)$, which is reproduced in
excellent agreement (Figs.\ \ref{fig:NaCl}e-h). This means that --at
least for the cases considered-- the enforced dissociation of NaCl can
be well described by a Markovian Langevin model. Since we simply added
an external force to an equilibrium dLE model, we have shown that the
external driving does not affect the validity of the friction and the
noise estimated at equilibrium conditions. Alternatively, we could
estimate the Langevin fields directly from the nonequilibrium data,
which is advantageous when dissipation-corrected targeted MD
simulations are available.\cite{Wolf20}

To demonstrate the limits of the Markovian approximation underlying
the dLE, we consider the velocity autocorrelation function $C_v$ shown
in Figs.\ \ref{fig:NaCl}i-l. It exhibits a rapid initial decay on a
$\sim 25\,$fs timescale, which is followed by damped oscillatory
features with a period of $\sim 120\,$fs, whose details depend on time
$t$ (see Ref.\ \onlinecite{Meyer20}). Similar as found for the
equilibrium dLE model,\cite{Lickert20} the Markovian LE correctly
obtains the initial decay of $C_v$, but underestimates the
oscillations. While the Markovian Langevin model fails to catch these
details on short timescales ($t \lesssim 0.1\,$ps), it reproduces
correctly the equilibrium and nonequilibrium ion dynamics on
timescales of 1 -- 1000 ps.

%
%
\subsection{Pressure-induced crystal nucleation of hard spheres}

As a potentially more challenging example, we next consider crystal
nucleation and growth processes in a compressed liquid of hard
spheres, which was recently studied as a test problem for the
nonstationary GLE.\cite{Kuhnhold19, Meyer21} Adopting 16384 hard
spheres with mass $m$ and diameter $\sigma$ and using a time step
$\delta t\!=\!\sqrt{m/(k_{\text{B}}T)}\sigma$, Meyer et
al.\cite{Meyer21} first equilibrated the system in a liquid state at a
volume fraction $\eta_0\!=\!0.45$. At time $t\!=\!0$ crystallization
is induced by impulsively compressing the system to a volume fraction
$\eta\!=\!0.54$ by rescaling the simulation box and all positions, and
$N_{\rm traj} \!=\!580$ trajectories of
$t_{\rm max}\!=\! 214 \, \delta t$ length were propagated. Using these
nonequilibrium trajectories as input data for a dLE, here we want to
study to what extent a Markovian Langevin model is able to describe
the complex cooperative processes underlying this weak first-order
phase transition.\cite{tenWolde95}

\begin{figure}[ht!]
\includegraphics[width=0.4\textwidth]{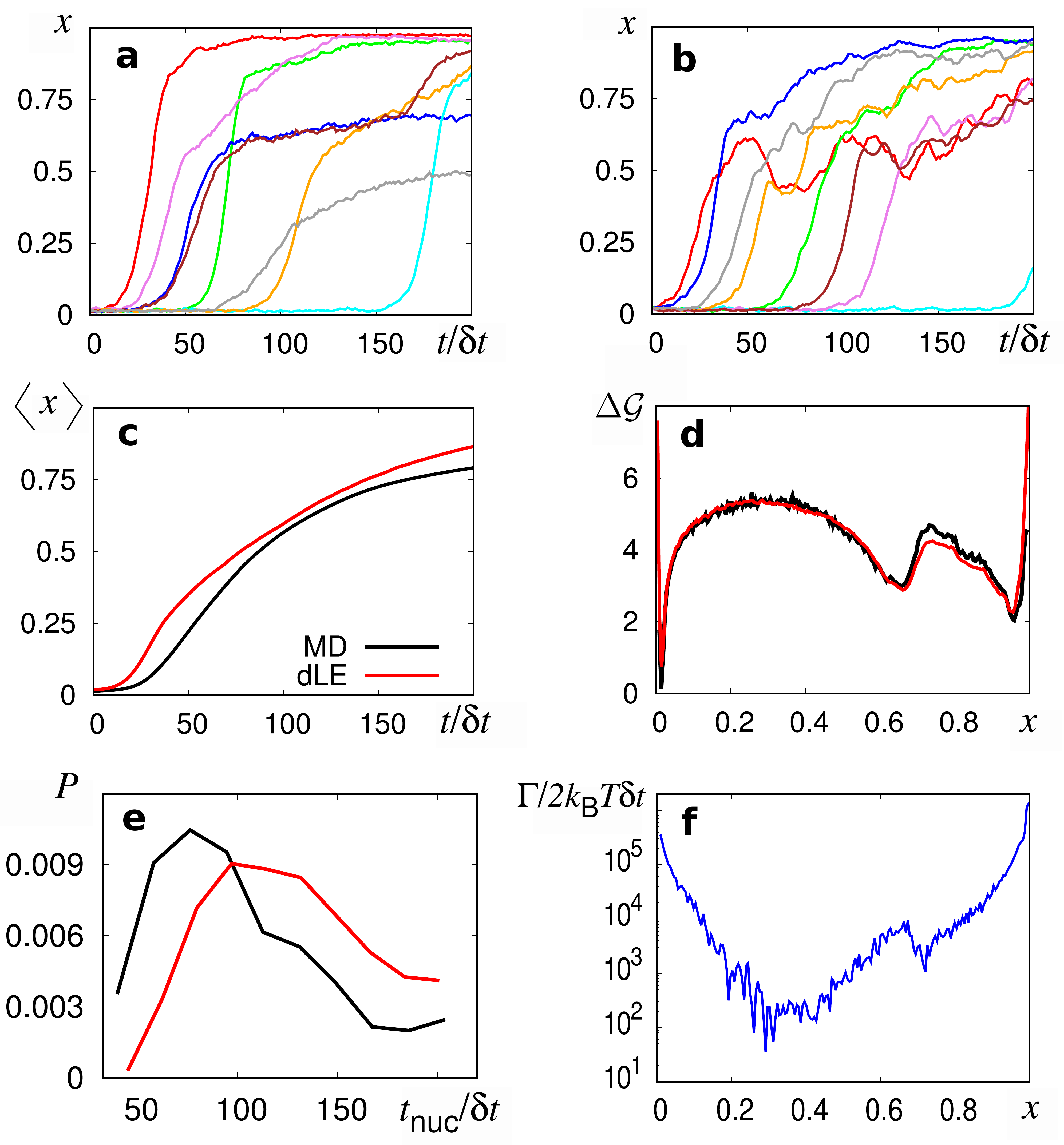}
\caption{
  Pressure-induced crystal nucleation of hard spheres, indicated by
  the percentage $x$ of particles that completed
  crystallization. Representative trajectories $x(t)$ obtained from
  (a) reference MD and (b) corresponding dLE simulations. Comparing MD
  and dLE results, shown are (c) the average percentage
  $\langle x(t)\rangle$, (d) the resulting energy landscape
  $\Delta{\cal G}(x)$, and (e) the distribution of nucleation times
  $t_{\rm nuc}$. Panel (f) shows the friction coefficient $\Gamma(x)$
  estimated by the dLE.}
\label{fig:nucleation} 
\end{figure}

As a one-dimensional reaction coordinate that accounts for the
crystallinity of the system, we choose the percentage $x$ of particles
that completed crystallization, which is readily calculated from the
$Q_6$ order parameter.\cite{Meyer21,tenWolde95} To get an overview of the
nucleation dynamics of the system, Fig.\ \ref{fig:nucleation}a shows
the time evolution $x(t)$ of eight sample trajectories. Due to the
mandatory preceding formation of nucleation seeds, the crystallization
process starts after some induction time, which is widely
distributed. The subsequent nucleation process is reflected in a rapid
sigmoidal-shaped rise of $x(t)$ on a timescale of $20\,\delta t$.  The
majority (357) of the trajectories rises to a value of
$x \! \gtrsim\!  0.8$ where they level off and slowly approach the
limiting value $x \!=\!1$. A smaller part (125) gets stuck at a value
of $x \! \sim\!  0.65$, reflecting the occurrence of crystal
defects. \cite{Meyer21,tenWolde95} The remaining trajectories start
too late to reach one of the two plateaus. Averaging over all
trajectories (Fig.\ \ref{fig:nucleation}c), the nucleation process is
found to occur on a timescale of $100$ $\delta t$.

Figure \ref{fig:nucleation}d shows the corresponding nonequilibrium
energy landscape $\Delta{\cal G} (x)$ defined in Eq.\ (\ref{eq:PMF}).
Here the minimum at $x\! \sim\! 0.014$ reflects the initial liquid
state of the system, and the minimum at $x\! \sim\!  0.65$ accounts
for the defective clusters. The minimum at $x\! \sim\!  0.95$
corresponds to (almost) completed crystals, and depends sensitively on
the finite length $t_{\rm max}$ of the nonequilibrium
trajectories. Increasing $t_{\rm max}$, more and more trajectories
reach the crystalline end state, which leads to a steep decrease of
the energy for $x\!\rightarrow\!1$. To define a nucleation time
$t_{\rm nuc}$, we choose $x\!=\!0.8$ as a threshold value for the
nucleation. Figure \ref{fig:nucleation}d shows the resulting
distribution of $t_{\rm nuc}$. Due to the relatively small sample size
(only 357 trajectories reach $x\!=\!0.8$), the distribution is given
at low resolution. It peaks at $t_{\rm nuc} = 76\,\delta t$ and yields
a mean nucleation time of $102\,\delta t$.

To construct a dLE model of the nonequilibrium MD data, we first
employ the noise test of Eq.\ (\ref{dLEnoise}) to verify that a dLE
using the data time step $\delta t$ results in a correct model of
white noise (see Fig.\ \SInucleationNoiseTest). Then $10^4$ dLE
trajectories of length $t_{\rm max}\!=\! 214 \, \delta t$ were
simulated, each run starting at $x=0.02$. Comparing the individual
trajectories of MD and dLE (Figs.\ \ref{fig:nucleation}a,b) as well as
their averages (Fig.\ \ref{fig:nucleation}c), we find that the dLE
qualitatively reproduces the induction time distribution and the
sigmoidal-shaped rise of $x(t)$. Moreover, the resulting
nonequilibrium energy landscape $\Delta{\cal G} (x)$ shown in Fig.\
\ref{fig:nucleation}d is found to be in excellent agreement with the
MD results. Unlike to the MD results, though, the dLE trajectories
mostly do not directly rise to $x \!\gtrsim\! 0.8$, but stay for some
time around $x \sim 0.65$ before they rise further. This deviation is
a consequence of the local estimation of the dLE fields [Eq.\
(\ref{knextneighav})], which does not distinguish between
crystallizing and defective trajectories.
As a consequence the resulting dLE distribution of nucleation times
$t_{\rm nuc}$ in Fig. \ref{fig:nucleation}e is somewhat shifted
compared to the MD distribution, which results in a mean dLE
nucleation time of $128\,\delta t$ (rather than $102\,\delta t$). Due
to the limited statistics of the MD data, though, this discrepancy
might change for better sampled data. For longer MD trajectories, for
example, we expect also longer nucleation times, since gradually all
non-defective trajectories crystallize.

Let us finally study the friction $\mG$ as a function of variable
$x$. Figure \ref{fig:nucleation}f shows that the friction is low at
the main barrier and high at the two main minima reflecting the liquid
and crystallized state. In line with previous results,\cite{Wolf20} we
find that the friction rises in particular at steep gradients of the
free energy landscape, where the system experiences large fluctuations
of the drift. At the onset of crystallization, these fluctuations may
arise from the large dissipation of excess energy after shock
compression. Since in other molecular systems (such as
peptides\cite{Schaudinnus15, Schaudinnus16} and NaCl above) the
friction is found to change only little (say, a factor 2 -- 10), this
variation by a factor of $\sim 10^3$ is remarkable.

%
%
\subsection{Global Langevin model from short trajectories}

A promising strategy for enhanced sampling is to combine massive
parallel computing of short MD trajectories to sample the free energy
landscape with a global dynamical model such as a Langevin or a Markov
state model\cite{Bowman09,Prinz11,Bowman13a} (MSM) to rebuild the
kinetics from the sampled data. As explained in Section \ref{sec:dLE},
this is possible because dLE fields as well as MSM transition matrices
are calculated locally.\cite{Rzepiela14} As short MD trajectories {\em
  per se} represent nonstationary data, there is again the question on
the convergence of the dLE model with respect to their number and
length (see Sec.\ \ref{sec:1dHierarch}). Moreover it is interesting to
compare the performance of a dLE to a MSM for the same data.

With this end in mind, we adopt the peptide Aib$_9$ as a
well-established model problem,\cite{Buchenberg15,Rzepiela14,
  Schaudinnus15,Perez18,Biswas18} for which long unbiased MD
data\cite{Buchenberg15} as well as an MSM built from numerous short MD
trajectories are available.\cite{Biswas18} That is, Buchenberg et
al.\cite{Buchenberg15} simulated $8\times 2\ \mu$s-long unbiased MD
trajectories at 320 K, and Biswas et al.\cite{Biswas18} employed
metadynamics\cite{Laio02,Bussi20} to generate initial conformations
for $\sim 7700$ unbiased trajectories of 10 ns length. All MD
simulations used the GROMACS program suite \cite{Abraham15} with the
GROMOS96 43a1 force field\cite{GROMOS96} and explicit chloroform
solvent.\cite{Tironi94}
For the analysis of the MD data, a principal component analysis on the
($\phi,\, \psi$) backbone dihedral angles (dPCA) of the five inner
residues of the peptide was performed,\cite{Altis08,Sittel17} whose
first five PCs account for 85\% of the total variance and exhibit
multipeaked distributions and a slow decay of the autocorrelation
function. Using these coordinates, robust density-based 
clustering\cite{Sittel16} was applied to identify metastable
conformational states.

\begin{figure}[ht!]
\includegraphics[width=0.49\textwidth]{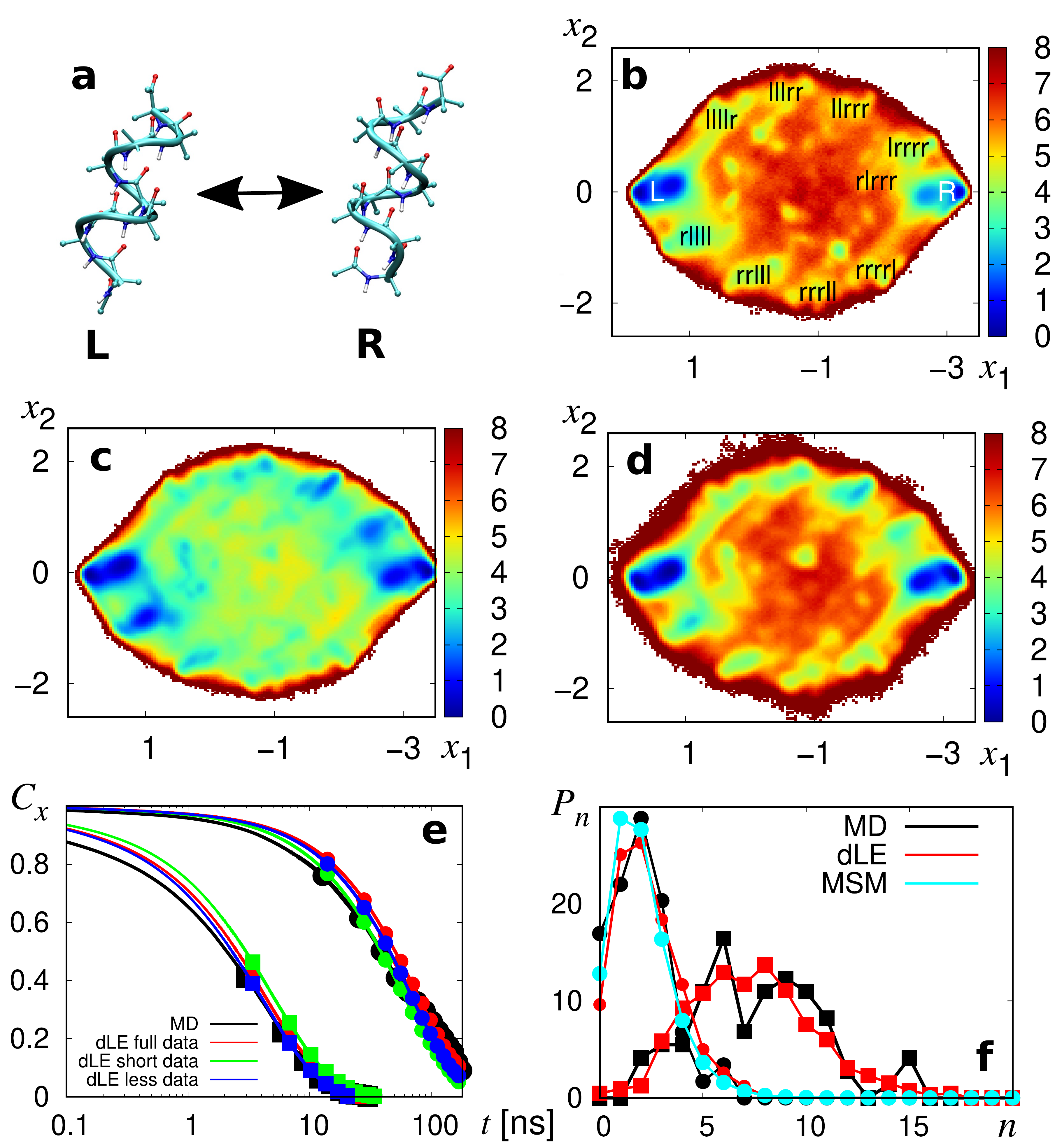}
\caption{
  (a) Conformational transitions of the peptide Aib$_9$ from a
  left-handed (L) to a right-handed (R) helix. Energy landscape
  $\Delta {\cal G}$ along the first two principal components of the
  system, obtained from (b) $8\!\times\! 2\ \mu$s-long unbiased MD
  simulations,\cite{Buchenberg15} (c) $\sim 7700$ 10\,ns-long unbiased
  trajectories seeded by metadynamics,\cite{Biswas18} and (d) the dLE
  model. (e) Autocorrelation functions of the first two principal
  components, as obtained from long MD runs and various dLE
  models. (f) Probability of pathways, which use at least one main
  intermediate state and exactly $n$ middle states, obtained from long
  unbiased MD simulations, \cite{Buchenberg15} the MSM of Biswas et
  al.\cite{Biswas18} and the dLE model. A time resolution of 1\,ns
  (\mycircle{black}) and 2 ps (\mysquare{black}) is used.}
\label{fig:Aib} 
\end{figure}

The conformational transitions of Aib$_9$ from a left- to a
right-handed helix (Fig.\ \ref{fig:Aib}a) exhibits complex structural
dynamics, which originates from various hierarchically coupled
dynamical processes on several timescales.\cite{Buchenberg15} Adopting
first the $8\times 2\ \mu$s-long unbiased MD data, Fig.\
\ref{fig:Aib}b shows the free energy landscape $\Delta G$ along the
first two principal components $x_1$ and $x_2$ of the dPCA. As
expected from the achirality of Aib$_9$, we find an overall symmetry
with respect to the first principal component $x_1$, where the two
main minima correspond to the all left-handed structure {L}
($x_1 \approx -2$) and the all right-handed structure {R}
($x_1 \approx 2$). Moreover, numerous metastable intermediate states
exist that constitute pathways from {L} to {R}. Containing a mixture
of right-handed (r) and left-handed (l) residues, the intermediate
states can uniquely be described by a product state of these
chiralities. Restricting ourselves to the five inner residues of
Aib$_9$, we obtain e.g., {L} = (lllll) and {R} = (rrrrr), as well as
(rllll) if all but residue 3 show left-handed conformations. The
resulting $2^5=32$ states provide a simple interpretation of the states
found by robust density-based clustering.

Judging from the prominent appearance of states \mbox{(lllll)}, (rlllll),
(rrlll),..., (rrrrr), (lrrrr), etc. in the free energy landscape, one
might expect that a sequential pathway along these states provides the
main way to get from {L} to {R} and back. These states are located at
the external border of the energy landscape, while other states such
as (rlrrr) are located inside the circle formed by the main states. This
finding was questioned by an MSM pathway analysis of short
implicit-solvent trajectories seeded by the MELD
protocol,\cite{Perez15} which revealed that transitions involving
lowly populated interior states nevertheless amounts to $\sim 40\%$ of the
overall flux.\cite{Perez18,Nagel20}
When we employ the $\sim 7700$ short unbiased trajectories seeded by
metadynamics,\cite{Biswas18} Fig.\ \ref{fig:Aib}c shows that the
simulations indeed cover the entire energetically accessible energy
landscape homogeneously. Since we deliberately started the short
trajectories all over the place including high-energy regions such as
barriers, however, this nonequilibrium energy landscape is not a good
representation of the Boltzmann weighted free energy landscape.

To obtain the correct free energy landscape from this nonequilibrium
data, we construct a dLE using
a time step $\delta t\!=\!2\,$ps which ensures an adequate resolution
of the dynamics. As described in Methods, we follow a recently
proposed strategy\cite{Lickert20} and rescale the friction and the
noise fields by an overall factor ($S \sim 2$), such that resulting
dLE model reproduces the initial decay of the position autocorrelation
functions (see Methods). To be able to use all $\sim 4\cdot 10^7$ data
points of the short trajectories, we applied a pre-averaging of the
data to $10^6$ points (see SI). In total $10$ dLE trajectories of
$20\,\mu$s length were simulated.
Figure \ref{fig:Aib}d shows the resulting free energy landscape, which
is quite similar to the landscape obtained for the long unbiased MD
simulations, except for a clearly better sampling of the lowly
populated intermediate states. Due to the rescaling, MD
and dLE also yield well matching autocorrelation functions of all five
principal components (Fig.\ \ref{fig:Aib}e and \SIAibACF), showing that the
first component decays on a $\sim 100\,$ns timescale, about an order
of magnitude slower than the other components. 

Adopting the state definition of Biswas et al.,\cite{Biswas18} we next
calculate mean waiting times of various transitions between the ten
most populated states and compared them to previous MSM results
\cite{Biswas18}, see Fig.\ \SIAibWaitingTimes. We find good overall
agreement of both approaches with an average deviation of 43\,\%. We
obtain a similar conformance (average deviation of 50\,\%) with the
waiting times from the long unbiased MD simulations, although this
comparison only makes sense for the well-sampled main transitions of
Aib$_9$.  As an example, Table \ref{tab:aib} compares the mean waiting
time obtained for transitions between the two main states R and
L. Being sampled 74 times by the long unbiased MD simulations, the MD
results may serve as a reference. In comparison, the dLE results (full
data) differ by 24 and 75\,\% for L$\rightarrow$R and R$\rightarrow$L,
respectively, while the MSM results differ by 18 and 16\,\%.

\begin{table}[h]
\centering
\begin{tabular}{l|l|l}
     & $\tau_{\text{L}\rightarrow\text{R}}$ [ns] & $\tau_{\text{R}\rightarrow\text{L}}$ [ns] \\
    \hline
    {MD} & $161\pm23$ & $80\pm12$ \\
    \hline
    {MSM} & $132\pm0.2$ & $67\pm0.1$ \\
    \hline
    {dLE (full data)} & $200\pm8$ & $140\pm6$ \\
    \hline
    {dLE (short data)} & $162\pm6$ & $113\pm4$ \\
    \hline
    {dLE (less data)} & $178\pm6$ & $132\pm5$ \\
\end{tabular}
\caption{
  Average waiting times of the L$\leftrightarrow$R transitions of
  Aib$_9$, obtained for $8\! \times\! 2\,\mu$s-long unbiased MD simulations,
  \cite{Buchenberg15} and for various models using short trajectories
  ($7700 \!\times\! 10\,$ns) of Biswas et al.\cite{Biswas18} as input
  data. Compared are a MSM,\cite{Biswas18} and various dLE models with
  full input data, and with shorter (4\,ns) and fewer (2000)
  trajectories. Errors are calculated as standard deviations of the mean.}
\label{tab:aib} 
\end{table}

To characterize various types of L$\rightarrow$R transition pathways,
Fig.\ \ref{fig:Aib}f shows the probability $P_n$ of pathways, which
use at least one main intermediate state and exactly $n$ middle
states. In nice agreement, long unbiased MD and dLE simulations
predict a unimodal distribution with a maximum for $n\approx 6$.
As the MSM requires a minimum lag time of 1\,ns, though, the resulting
distribution is shifted towards shorter pathways with a peak at 
$n\approx 2.5$. Similar results are also obtained from MD and dLE,
when a time resolution of 1\,ns is used. Hence we find that, allowing
for shorter time step than an MSM, the dLE may possibly provide a
better time resolution of the dynamics.

To study the performance of the dLE for limited input data, we run two
more sets of dLE simulations, where one set uses shorter
($7700 \times 4\,$ns) and the other one fewer ($2000 \times 10\,$ns)
input trajectories. When we compare the resulting autocorrelation
functions (Fig.\ \ref{fig:Aib}e and \SIAibACF) and the
L$\leftrightarrow$R waiting times (Table \ref{tab:aib}) to results
obtained for full input data, we again find that reduced input data in
general leads to faster dynamics. This is particularly the case for
shorter input data, which lead to a $\sim 20\,$\% decrease of the
L$\leftrightarrow$R waiting times.
Similarly as found for the one-dimensional system in Sec.\
\ref{sec:1dHierarch}, the five-dimensional dLE model for AiB$_9$ is
shown to robustly predict the overall kinetics from nonequilibrium
input data.

%
%
\section{Conclusions}

To extend the application of the data-driven Langevin equation (dLE)
to the nonequilibrium regime, we have introduced various modifications
of the original approach.\cite{Hegger09,Schaudinnus13, Schaudinnus15,
  Schaudinnus16,Lickert20} As a central result, we have shown that
nonequilibrium conditions mainly affect the deterministic drift term
$\vf$ of the dLE (\ref{dLE}). Compared to a system at thermal
equilibrium which performs a Boltzmann sampling of the free energy
landscape ($\vf = - \partial_{\vx} \Delta G(\vx)$), nonequilibrium
processes sample the biased energy landscape
\begin{equation} \label{eq:PMF3}
\Delta{\cal G} (\vx, t) =-k_{\rm B}T \ln {\cal P}(\vx) + V_{\rm ext}(\vx,t).
\end{equation}
Here the first term accounts for the finite sampling in nonequilibrium
simulations. For example, describing a relaxation process (Fig.\
\ref{fig:neq}a) with trajectories of length $t_{\rm max}$,
${\cal P}(\vx)$ reflects the conformational distribution sampled
within $0 \le t \le t_{\rm max}$. As a general result, we have found
that too short input data bias the prediction towards too fast
dynamics.
The second term represents a time-dependent potential that accounts
for the driving of the system. Considering, for example, an atomic
force microscopy experiment that enforces the dissociation of a
molecule (Fig.\ \ref{fig:neq}b), $V_{\rm ext}$ describes the gradual
pulling of the system along the energy landscape. The biased energy
landscape $\Delta{\cal G}(\vx, t)$ thus corresponds to the total
deterministic potential experienced by the system at time $t$.

The dLE modeling of relaxation processes does not require additional
assumptions compared to an equilibrium dLE. Indeed, for all considered
examples of nonstationary preparation including hierarchical dynamics
(Fig.\ \ref{fig:1dmodel}), crystal nucleation (Fig.\
\ref{fig:nucleation}), and peptide conformational dynamics (Fig.\
\ref{fig:Aib}), the resulting dLE model was shown to reproduce the
reference data convincingly. On the other hand, it is in general an
approximation that an external force leaves the friction $\mG$ and the
noise $\mK$ describing the system's coupling to the bath
unchanged. This assumption was shown to hold in the linear-response
regime, which is expected to be valid for most biomolecular
applications. For example, we found linear response behavior for the
enforced dissociation of NaCl (Fig.\ \ref{fig:NaCl}) up to
extremely high pulling velocities.\cite{Wolf18}

Finally we have compared the dLE approach to the nonstationary GLE of
Meyer et al.\cite{Meyer17,Meyer19} Studying crystal nucleation and
enforced dissociation of NaCl, the GLE analyses\cite{Meyer20,Meyer21}
resulted in slowly decaying memory kernels, which appears to question
the applicability of a Markovian Langevin model such as the dLE. A
theoretical analysis revealed that the LE (\ref{eq:LE}) and the GLE
(\ref{eq:nsGLE}) are based on different partitionings between system
and bath, such that the former may well result in a Markovian system,
although the latter exhibits non-Markovian behavior. The formulations
thus provide a complementary description of nonequilibrium
processes. The nonstationary GLE represents in principle an exact
formulation, however, its memory kernel and the associated colored
noise in general may be difficult to determine and interpret, in
particular when we want to go beyond a one-dimensional phase space
variable. While a Markovian LE model relies on a suitably chosen
system coordinate, it represents a physically intuitive model, where
the biased energy landscape $\Delta{\cal G}$ accounts for the main
features of the considered process, and the friction $\mG$ and the
noise $\mK$ describe in a transparent way the effects of the
bath. Moreover, the dLE algorithm allows us to use multidimensional
coordinates $\vx$ and fields $\vf$, $\mG$ and $\mK$ that depend on
$\vx$, which greatly adds to the versatility of the method.

%
%
\subsection*{Supporting Information} \vspace{-3mm}
Details on the pre-averaging method, friction fields of the
hierarchical model, noise checks of the nucleation model, as well as
noise checks, autocorrelation functions, and mean waiting times
of Aib$_9$.

The dLE algorithm as described in Ref.\ \onlinecite{Schaudinnus16} including 
pre-averaging and rescaling is freely available at
https://github.com/moldyn.

All data shown are available from the corresponding author upon
reasonable request. 

\subsection*{Acknowledgment}\vspace{-3mm}
We thank Tanja Schilling and her group for helpful discussions and for
sharing their trajectories of the crystal nucleation problem, as well as
Matthias Post and Mithun Biswas for numerous instructive discussions.
This work has been supported by the DFG via the Research Unit FOR5099
``Reducing complexity of nonequilibrium systems'', by the bwUniCluster
computing initiative, the High Performance and Cloud Computing Group
at the Zentrum f\"ur Datenverarbeitung of the University of
T\"ubingen, the state of Baden-W\"urttemberg through bwHPC and the DFG
through grant No. INST 37/935-1 FUGG.

%
%


\begin{thebibliography}{10}

\bibitem{Berendsen07}
H.~J.~C. Berendsen,
\newblock {\em Simulating the Physical World},
\newblock Cambridge University Press, Cambridge, 2007.

\bibitem{Rohrdanz13}
M.~A. Rohrdanz, W.~Zheng, and C.~Clementi,
\newblock Discovering mountain passes via torchlight: Methods for the
  definition of reaction coordinates and pathways in complex macromolecular
  reactions,
\newblock Annu. Rev. Phys. Chem. {\bf 64}, 295 (2013).

\bibitem{Peters16}
B.~Peters,
\newblock Reaction coordinates and mechanistic hypothesis tests,
\newblock Annu. Rev. Phys. Chem. {\bf 67}, 669 (2016).

\bibitem{Sittel18}
F.~Sittel and G.~Stock,
\newblock Perspective: Identification of collective coordinates and metastable
  states of protein dynamics,
\newblock J. Chem. Phys. {\bf 149}, 150901 (2018).

\bibitem{Grabert80}
H.~Grabert, P.~{H\"anggi}, and P.~Talkner,
\newblock Microdynamics and nonlinear stochastic processes of gross variables,
\newblock J. Stat. Phys. {\bf 22}, 537 – 552 (1980).

\bibitem{Kubo85}
R.~Kubo, M.~Toda, and N.~Hashitsume,
\newblock {\em Statistical {P}hysics~II. {N}onequilibrium {S}tatistical
  {M}echanics},
\newblock Springer, Berlin, 1985.

\bibitem{Zwanzig01}
R.~Zwanzig,
\newblock {\em Nonequilibrium Statistical Mechanics},
\newblock Oxford University, Oxford, 2001.

\bibitem{Hernandez99}
R.~Hernandez,
\newblock The projection of a mechanical system onto the irreversible
  generalized {Langevin} equation,
\newblock J. Chem. Phys. {\bf 111}, 7701 (1999).

\bibitem{McPhie01}
M.~McPhie, P.~Daivis, I.~Snook, J.~Ennis, and D.~Evans,
\newblock Generalized {Langevin} equation for nonequilibrium systems,
\newblock Physica A {\bf 299}, 412 (2001).

\bibitem{Micheletti08}
C.~Micheletti, G.~Bussi, and A.~Laio,
\newblock Optimal {Langevin} modeling of out-of-equilibrium molecular dynamics
  simulations,
\newblock J. Chem. Phys. {\bf 129}, 074105 (2008).

\bibitem{Kawai11}
S.~Kawai and T.~Komatsuzaki,
\newblock Derivation of the generalized {Langevin} equation in nonstationary
  environments,
\newblock J. Chem. Phys. {\bf 134}, 114523 (2011).

\bibitem{Meyer17}
H.~Meyer, T.~Voigtmann, and T.~Schilling,
\newblock {On the non-stationary generalized Langevin equation},
\newblock J. Chem. Phys. {\bf 147}, 214110 (2017).

\bibitem{Meyer19}
H.~Meyer, T.~Voigtmann, and T.~Schilling,
\newblock On the dynamics of reaction coordinates in classical, time-dependent,
  many-body processes,
\newblock J. Chem. Phys. {\bf 150}, 174118 (2019).

\bibitem{Cui18}
B.~Cui and A.~Zaccone,
\newblock Generalized {Langevin} equation and fluctuation-dissipation theorem
  for particle-bath systems in external oscillating fields,
\newblock Phys. Rev. E {\bf 97}, 060102 (2018).

\bibitem{Zwanzig73}
R.~Zwanzig,
\newblock Nonlinear generalized {Langevin} equations,
\newblock J. Stat. Phys. {\bf 9}, 215  (1973).

\bibitem{Straub87}
J.~E. Straub, M.~Borkovec, and B.~J. Berne,
\newblock Calculation of dynamic friction on intramolecular degrees of freedom,
\newblock J. Phys. Chem. {\bf 91}, 4995  (1987).

\bibitem{Hummer05}
G.~Hummer,
\newblock Position-dependent diffusion coefficients and free energies from
  {Bayesian} analysis of equilibrium and replica molecular dynamics
  simulations,
\newblock New J. Phys. {\bf 7}, 34 (2005).

\bibitem{Best06}
R.~B. Best and G.~Hummer,
\newblock Diffusive model of protein folding dynamics with {Kramers} turnover
  in rate,
\newblock Phys. Rev. Lett. {\bf 96}, 228104 (2006).

\bibitem{Lange06b}
O.~F. Lange and H.~{Grubm\"uller},
\newblock Collective {Langevin} dynamics of conformational motions in proteins,
\newblock J. Chem. Phys. {\bf 124}, 214903 (2006).

\bibitem{Hinczewski10}
M.~Hinczewski, Y.~von Hansen, J.~Dzubiella, and R.~R. Netz,
\newblock How the diffusivity profile reduces the arbitrariness of protein
  folding free energies,
\newblock J. Chem. Phys. {\bf 132}, 245103 (2010).

\bibitem{Perez-Villa18}
A.~Perez-Villa and F.~Pietrucci,
\newblock Free energy, friction, and mass profiles from short molecular
  dynamics trajectories,
\newblock arXiv {\bf 1810.00713} (2018).

\bibitem{Wolf18}
S.~Wolf and G.~Stock,
\newblock Targeted molecular dynamics calculations of free energy profiles
  using a nonequilibrium friction correction,
\newblock J. Chem. Theory Comput. {\bf 14}, 6175  (2018).

\bibitem{Paul19}
F.~Paul, H.~Wu, M.~Vossel, B.~L. {de Groot}, and F.~No{\'e},
\newblock Identification of kinetic order parameters for non-equilibrium
  dynamics,
\newblock J. Chem. Phys. {\bf 150}, 164120 (2019).

\bibitem{Hegger09}
R.~Hegger and G.~Stock,
\newblock Multidimensional {Langevin} modeling of biomolecular dynamics,
\newblock J. Chem. Phys. {\bf 130}, 034106 (2009).

\bibitem{Schaudinnus13}
N.~Schaudinnus, A.~J. Rzepiela, R.~Hegger, and G.~Stock,
\newblock Data driven {Langevin} modeling of biomolecular dynamics,
\newblock J. Chem. Phys. {\bf 138}, 204106 (2013).

\bibitem{Schaudinnus15}
N.~Schaudinnus, B.~Bastian, R.~Hegger, and G.~Stock,
\newblock Multidimensional {Langevin} modeling of nonoverdamped dynamics,
\newblock Phys. Rev. Lett. {\bf 115}, 050602 (2015).

\bibitem{Schaudinnus16}
N.~Schaudinnus, B.~Lickert, M.~Biswas, and G.~Stock,
\newblock Global {Langevin} model of multidimensional biomolecular dynamics,
\newblock J. Chem. Phys. {\bf 145}, 184114 (2016).

\bibitem{Lickert20}
B.~Lickert and G.~Stock,
\newblock Modeling {non-Markovian} data using {Markov} state and {Langevin}
  models,
\newblock J. Chem. Phys. {\bf 153}, 244112 (2020).

\bibitem{Meyer20}
H.~Meyer, S.~Wolf, G.~Stock, and T.~Schilling,
\newblock A numerical procedure to evaluate memory effects in non-equilibrium
  coarse-grained models,
\newblock Adv. Theory Simul. {\bf 111}, 2000197 (2020).

\bibitem{Meyer21}
H.~Meyer, F.~Glatzel, W.~{W\"ohler}, and T.~Schilling,
\newblock Evaluation of memory effects at phase transitions and during
  relaxation processes,
\newblock Phys. Rev. E {\bf 103}, 022102 (2021).

\bibitem{Rzepiela14}
A.~J. Rzepiela, N.~Schaudinnus, S.~Buchenberg, R.~Hegger, and G.~Stock,
\newblock Communication: Microsecond peptide dynamics from nanosecond
  trajectories: A {Langevin} approach,
\newblock J. Chem. Phys. {\bf 141}, 241102 (2014).

\bibitem{Buchenberg15}
S.~Buchenberg, N.~Schaudinnus, and G.~Stock,
\newblock Hierarchical biomolecular dynamics: Picosecond hydrogen bonding
  regulates microsecond conformational transitions,
\newblock J. Chem. Theory Comput. {\bf 11}, 1330 (2015).

\bibitem{Biswas18}
M.~Biswas, B.~Lickert, and G.~Stock,
\newblock Metadynamics enhanced {Markov} modeling: Protein dynamics from short
  trajectories,
\newblock J. Phys. Chem. B {\bf 122}, 5508  (2018).

\bibitem{Grassberger90}
P.~Grassberger,
\newblock An optimized box-assisted algorithm for fractal dimensions,
\newblock Phys. Lett. A {\bf 148}, 63 (1990).

\bibitem{Post19}
M.~Post, S.~Wolf, and G.~Stock,
\newblock Principal component analysis of nonequilibrium molecular dynamics
  simulations,
\newblock J. Chem. Phys. {\bf 150}, 204110 (2019).

\bibitem{Daldrop17}
J.~O. Daldrop, B.~G. Kowalik, and R.~R. Netz,
\newblock External potential modifies friction of molecular solutes in water,
\newblock Phys. Rev. X {\bf 7}, 041065 (2017).

\bibitem{Chandler87}
D.~Chandler,
\newblock {\em Introduction to Modern Statistical Mechanics},
\newblock Oxford University, Oxford, 1987.

\bibitem{Grubmueller96}
H.~{Grubm\"uller}, B.~Heymann, and P.~Tavan,
\newblock Ligand binding: Molecular mechanics calculation of the
  streptavidin-biotin rupture force,
\newblock Science {\bf 271}, 997 (1996).

\bibitem{Isralewitz01}
B.~Isralewitz, M.~Gao, and K.~Schulten,
\newblock {Steered molecular dynamics and mechanical functions of proteins},
\newblock Curr. Opin. Struct. Biol. {\bf 11}, 224 (2001).

\bibitem{Park04}
S.~Park and K.~Schulten,
\newblock Calculating potentials of mean force from steered molecular dynamics
  simulations,
\newblock J. Chem. Phys. {\bf 120}, 5946 (2004).

\bibitem{Frauenfelder91}
H.~Frauenfelder, S.~Sligar, and P.~Wolynes,
\newblock The energy landscapes and motions of proteins,
\newblock Science {\bf 254}, 1598 (1991).

\bibitem{Henzler-Wildman07}
K.~Henzler-Wildman and D.~Kern,
\newblock Dynamic personalities of proteins,
\newblock Nature (London) {\bf 450}, 964  (2007).

\bibitem{Buchenberg17}
S.~Buchenberg, F.~Sittel, and G.~Stock,
\newblock Time-resolved observation of protein allosteric communication,
\newblock Proc. Natl. Acad. Sci. USA {\bf 114}, E6804 (2017).

\bibitem{Bozovic20}
O.~Bozovic, C.~Zanobini, A.~Gulzar, B.~Jankovic, D.~Buhrke, M.~Post, S.~Wolf,
  G.~Stock, and P.~Hamm,
\newblock Real-time observation of ligand-induced allosteric transitions in a
  {PDZ} domain,
\newblock Proc. Natl. Acad. Sci. USA {\bf 117}, 26031  (2020).

\bibitem{Gernert14}
R.~Gernert, C.~Emary, and S.~Klapp,
\newblock Waiting time distribution for continous stochastic systems,
\newblock Phys. Rev. E {\bf 90}, 062115 (2014).

\bibitem{Wolf20}
S.~Wolf, B.~Lickert, S.~Bray, and G.~Stock,
\newblock Multisecond ligand dissociation dynamics from atomistic simulations,
\newblock Nat. Commun. {\bf 11}, 2918 (2020).

\bibitem{Abraham15}
M.~J. Abraham, T.~Murtola, R.~Schulz, S.~Pall, J.~C. Smith, B.~Hess, and
  E.~Lindahl,
\newblock Gromacs: High performance molecular simulations through multi-level
  parallelism from laptops to supercomputers,
\newblock SoftwareX {\bf 1}, 19  (2015).

\bibitem{Hornak06}
V.~Hornak, R.~Abel, A.~Okur, B.~Strockbine, A.~Roitberg, and C.~Simmerling,
\newblock Comparison of multiple {Amber} force fields and development of
  improved protein backbone parameters,
\newblock Proteins {\bf 65}, 712 (2006).

\bibitem{Best09}
R.~B. Best and G.~Hummer,
\newblock Optimized molecular dynamics force fields applied to the helix-coil
  transition of polypeptides,
\newblock J. Phys. Chem. B {\bf 113}, 9004 (2009).

\bibitem{Jorgensen83}
W.~L. Jorgensen, J.~Chandrasekhar, J.~D. Madura, R.~W. Impey, and M.~Klein,
\newblock Comparison of simple potential functions for simulating liquid water,
\newblock J. Chem. Phys. {\bf 79}, 926 (1983).

\bibitem{Mullen14}
R.~G. Mullen, J.-E. Shea, and B.~Peters,
\newblock {Transmission Coefficients, Committors, and Solvent Coordinates in
  Ion-Pair Dissociation},
\newblock J. Chem. Theory Comput. {\bf 10}, 659 (2014).

\bibitem{Kuhnhold19}
A.~Kuhnhold, H.~Meyer, G.~Amati, P.~Pelagejcev, and T.~Schilling,
\newblock Derivation of an exact, nonequilibrium framework for nucleation:
  Nucleation is a priori neither diffusive nor markovian,
\newblock Phys. Rev. E {\bf 100}, 052140 (2019).

\bibitem{tenWolde95}
P.~R. ten Wolde, M.~J. Ruiz-Montero, and D.~Frenkel,
\newblock Numerical evidence for bcc ordering at the surface of a critical fcc
  nucleus,
\newblock Phys. Rev. Lett. {\bf 75}, 2714 (1995).

\bibitem{Bowman09}
G.~R. Bowman, K.~A. Beauchamp, G.~Boxer, and V.~S. Pande,
\newblock Progress and challenges in the automated construction of {Markov}
  state models for full protein systems,
\newblock J. Chem. Phys. {\bf 131}, 124101 (2009).

\bibitem{Prinz11}
J.-H. Prinz, H.~Wu, M.~Sarich, B.~Keller, M.~Senne, M.~Held, J.~D. Chodera,
  C.~{Sch\"utte}, and F.~No{\'e},
\newblock Markov models of molecular kinetics: generation and validation,
\newblock J. Chem. Phys. {\bf 134}, 174105 (2011).

\bibitem{Bowman13a}
G.~R. Bowman, V.~S. Pande, and F.~No{\'e},
\newblock {\em An Introduction to Markov State Models},
\newblock Springer, Heidelberg, 2013.

\bibitem{Perez18}
A.~Perez, F.~Sittel, G.~Stock, and K.~Dill,
\newblock Meld-path efficiently computes conformational transitions, including
  multiple and diverse paths,
\newblock J. Chem. Theory Comput. {\bf 14}, 2109  (2018).

\bibitem{Laio02}
A.~Laio and M.~Parrinello,
\newblock Escaping free-energy minima,
\newblock Proc. Natl. Acad. Sci. USA {\bf 99}, 12562  (2002).

\bibitem{Bussi20}
G.~Bussi, A.~Laio, and P.~Tiwary,
\newblock {\em Metadynamics: A Unified Framework for Accelerating Rare Events
  and Sampling Thermodynamics and Kinetics}, pages 565--595,
\newblock Springer International Publishing, Cham, 2020.

\bibitem{GROMOS96}
W.~F. {van Gunsteren}, S.~R. Billeter, A.~A. Eising, P.~H. {H\"unenberger},
  P.~{Kr\"uger}, A.~E. Mark, W.~R.~P. Scott, and I.~G. Tironi,
\newblock {\em Biomolecular Simulation: The GROMOS96 Manual and User Guide},
\newblock {Vdf Hochschulverlag AG an der ETH Z\"urich}, Z\"urich, 1996.

\bibitem{Tironi94}
I.~G. Tironi and W.~F. {van Gunsteren},
\newblock A molecular dynamics simulation study of chloroform,
\newblock Mol. Phys. {\bf 83}, 381 (1994).

\bibitem{Altis08}
A.~Altis, M.~Otten, P.~H. Nguyen, R.~Hegger, and G.~Stock,
\newblock Construction of the free energy landscape of biomolecules via
  dihedral angle principal component analysis,
\newblock J. Chem. Phys. {\bf 128}, 245102 (2008).

\bibitem{Sittel17}
F.~Sittel, T.~Filk, and G.~Stock,
\newblock Principal component analysis on a torus: Theory and application to
  protein dynamics,
\newblock J. Chem. Phys. {\bf 147}, 244101 (2017).

\bibitem{Sittel16}
F.~Sittel and G.~Stock,
\newblock Robust density-based clustering to identify metastable conformational
  states of proteins,
\newblock J. Chem. Theory Comput. {\bf 12}, 2426 (2016).

\bibitem{Perez15}
A.~Perez, J.~L. MacCallum, and K.~Dill,
\newblock {Accelerating molecular simulations of proteins using Bayesian
  inference on weak information.},
\newblock Proc. Natl. Acad. Sci. USA {\bf 112}, 11846 (2015).

\bibitem{Nagel20}
D.~Nagel, A.~Weber, and G.~Stock,
\newblock {MSMPathfinder}: Identification of pathways in {Markov} state models,
\newblock J. Chem. Theory Comput. {\bf 16}, 7874  (2020).

\end{thebibliography}


\end{document}